\documentclass[preprint]{aastex63}
\usepackage{amsmath}
\usepackage{apjfonts}
\usepackage{IEEEtrantools}
\usepackage{url}
\usepackage{color}

\received{October 11, 2023}
\revised{December 12, 2023}
\accepted{January 8, 2024}
\submitjournal{PASP}

\shorttitle{Mitigating Charge Migration in H2RG data from JWST}
\shortauthors{Goudfrooij et al.}

\begin{document}

\title{An Algorithm to Mitigate Charge Migration Effects in
  Data from the Near Infrared Imager and Slitless
  Spectrograph on the James Webb Space Telescope\footnote{This work is
  based on observations made with the NASA/ESA/CSA James
  Webb Space Telescope. The data were obtained from the Mikulski
  Archive for Space Telescopes at the Space Telescope Science
  Institute, which is operated by the Association of Universities for
  Research in Astronomy, Inc., under NASA contract NAS 5-03127 for
  JWST. These observations are associated with programs \#\,1083, 1093,
  1094, and 1096.}} 

\correspondingauthor{Paul Goudfrooij}
\email{goudfroo@stsci.edu}

\author[0000-0002-5728-1427]{Paul Goudfrooij}
\affiliation{Space Telescope Science Institute, 3700 San Martin
  Drive, Baltimore, MD 21218, USA}

\author{David Grumm}
\affiliation{Space Telescope Science Institute, 3700 San Martin
  Drive, Baltimore, MD 21218, USA}

\author[0000-0002-3824-8832]{Kevin Volk}
\affiliation{Space Telescope Science Institute, 3700 San Martin
  Drive, Baltimore, MD 21218, USA}
\affiliation{Affiliated to the Canadian Space Agency}

\author{Howard Bushouse}
\affiliation{Space Telescope Science Institute, 3700 San Martin
  Drive, Baltimore, MD 21218, USA}

\begin{abstract}
We present an algorithm that mitigates the effects of charge migration due to
the ``brighter-fatter effect'' (BFE) that occurs for highly illuminated
stars in the Teledyne HAWAII-2RG detectors used in the 
NIRCam, NIRISS, and NIRSpec science instruments aboard the James Webb
Space Telescope (JWST). The impact of this effect is most significant for 
photometry and spectrophotometry of bright stars in data for which the
point spread function (PSF) is undersampled, which is the case for
several observing modes of the NIRISS instrument.
The main impact of BFE to NIRISS data is incorrect count rate determinations
for pixels in the central regions of PSFs of bright stars due to jump
detections that are caused by charge migration from peak pixels to surrounding
pixels. 
The effect is especially significant for bright compact sources in resampled,
distortion-free images produced by the {\it drizzle} algorithm:
quantitatively, apparent flux losses of $> 50$\% can occur in such images due to
BFE.  
We describe the algorithm of the ``{\tt charge\_migration}'' mitigation step that
has been implemented in version 10.0 of the operational JWST
calibration pipeline as of Dec 5, 2023.  We illustrate the impact of
this step in terms of the resulting improvements of the precision of
imaging photometry of point sources. The algorithm renders the effects
of BFE on photometry and surface brightness measurements to stay within 1\%.    
\end{abstract}


\section{Introduction} \label{sec:intro}
The Near InfraRed Imager and Slitless Spectrograph (NIRISS; \citealt{doyon+23})
on board the \emph{James Webb Space Telescope} \citep[\emph{JWST,}][]{gardner+23}
has four observing modes:
(1) aperture masking interferometry \citep[AMI;][]{siva+23}, (2) direct imaging,
(3) single-object slitless spectroscopy \citep[SOSS;][]{albert+23}, and (4) wide 
field slitless spectroscopy \citep[WFSS;][]{willott+22}. NIRISS uses a single
{\sc Hawaii-2RG} (H2RG) HgCdTe array manufactured by Teledyne Imaging Systems as
its detector, covering a useful wavelength range out to 5 $\mu$m. It features a
pixel size of 0\farcs0656, for which the JWST point spread function (PSF) is
critically sampled at a wavelength $\lambda \sim$\, 4 $\mu$m. 

The brighter-fatter effect (BFE) is a non-linear process that blurs the
intensity distribution of brighter sources on the detector to a larger extent
than it does for fainter sources. BFE was first observed and characterized in
charge coupled devices (CCDs) of several instruments such as Euclid
\citep{niemi+15}, the Dark Energy Camera  \citep{gruen+15} and the LSST/Rubin
telescope \citep{lage+17}. In CCDs, the effect is due to changes in the electric
field geometry within detector pixels as photoelectrons accumulate within the
pixel potential well \citep[e.g.,][]{antilogus+14}. Further
accumulation of photoelectrons is progressively hindered by the
increasing transverse electric field which repulses additional
incoming photoelectrons to neighboring pixels \cite[see also][]{hirachoi20}.
In NIR detectors such as those on JWST, where photo-generated charges are
collected in a depletion region generated at a p-n diode at the detector layer
which induces a change of voltage that is read using non-destructive sampling,
the physical reason to expect a BFE is different: as charge accumulates in a
pixel, the substrate voltage changes and the local depletion region shrinks. If
it shrinks significantly relative to that of a neighboring pixel, then new
charge generated in that area has a larger probability to get collected in the
neighboring pixel (with larger depletion region). The effect has been reported
in ground testing of a H2RG near-infrared detector for Euclid
(\citealt{plazas+17,plazas+18}; see also \citealt{zengilowski+21}). The
latter differ from the H2RG detectors used on 
JWST in terms of wavelength coverage, with the Euclid devices having a HgCdTe
cutoff of 2.3 $\mu$m, while most JWST devices cut off at $\sim$\,5.2 $\mu$m.
As such, a study of the impact of the BFE on science with H2RG detectors on JWST
seems warranted.

The BFE is not the only effect that involves nearest-neighbor
  interactions between H2RG detector pixels. Infrared detectors
  also suffer from electronic cross-talk due to
  capacitive coupling between neighboring pixels, an effect
  usually referred to as inter-pixel capacitance (IPC). Although
  the main effect of IPC is signal-independent, it can have a
  non-linear component that is signal-dependent \citep[NL-IPC; see,
e.g.,][and references therein]{donlon+18}). While the data and
  analysis used in the current paper formally do not allow one to
  separate the effects of NL-IPC and BFE, we note that 
  \citet{hirachoi20} introduced a framework to connect the
  cross-correlation signal of different flat field time samples to
  different non-linear detector behaviors. This formalism was applied 
  to a large dataset of flat field exposures with long ramps taken with a
  development H4RG detector for the \emph{WFIRST} (now \emph{Roman}) mission by 
  \citet{choihira20} and \citet{freudenburg+20}. In each of several different  
  tests, they found that the BFE dominated over the NL-IPC. In this
  paper, we assume that the signal-dependent effect of charge
  transfer to neighboring pixels in JWST H2RG devices is due to the BFE. 
    
An important feature of the BFE is that the magnitude of its effect scales with
pixel-to-pixel contrast: the larger the contrast between the charge
accumulated in neighboring pixels, the more efficient is the transfer of charge 
from the brightest pixel to its neighbors. As such, the effect is strongest for
bright point sources, especially in observing modes for which the point spread
function (PSF) is undersampled by the detector pixels. Severe PSF
undersampling with JWST occurs for three NIRISS observing modes. This includes
NIRISS Imaging and WFSS using filter passbands at wavelengths $\la 2 \;\mu$m,
for which the PSF is undersampled by factors $\ga$ 2, and AMI
observations with the non-redundant aperture mask for which the 
spatial resolution is given by the Michelson criterion ($\delta \theta = 0.5
\lambda/D$), a resolution roughly twice as high as for regular direct
imaging \citep[e.g.,][]{siva+23}. 

The BFE is particularly problematic for projects that rely on PSF
modeling to provide the highest possible precision in photometric, astrometric,
or morphological measurements. Good examples in terms of JWST science are
PSF-fitting photometry of point sources for studies of resolved stellar
populations and/or high-precision proper motion measurements of sources that are
either too faint for GAIA or in regions that are too crowded to be resolved by
GAIA \citep[e.g.,][]{libralato+23}, or cosmological studies of weak lensing and
cosmic shear \citep[e.g.,][]{amara+08,amara+10}.
The issue is that a systematic misrepresentation of the PSF when
measured from profiles of bright stars (to reach the necessary
signal-to-noise ratio) biases the resulting brightnesses of fainter  
stars, or shape measurements of galaxies, to levels that can significantly limit
the possible science goals. 
An additional issue caused by the BFE that specifically affects data that is
read out using non-destructive reads (i.e., NIR and mid-IR data) is that the BFE
changes the effective count rate during integration ramps, rendering the ramp
non-linear. For pixels with intrinsically high count rates, this causes false
positives in outlier detection schemes during detector-level data processing.  
In the absence of a BFE mitigation algorithm, this causes problems for point
source photometry when multiple dithered images are combined and resampled onto
a common distortion-free pixel grid using the drizzle algorithm \citep{fruhoo02}
in conjunction with the common weighting method of inverse variance mapping
\citep[IVM; see][]{caser+00}. This is discussed in detail in Section~\ref{s:impact}. 

In this paper we describe the effects of BFE on NIRISS data and its impact on
science, and we introduce an algorithm that mitigates these effects and was
recently implemented as a new step in the JWST Calibration Pipeline. 

\section{Data Processing} \label{s:data}
Before describing examples of the impact of BFE on NIRISS data, we
briefly review the relevant processing steps in the JWST Calibration
Pipeline \citep[see][]{bushouse+23}. 

\subsection{Overview of JWST Pipeline Processing of H2RG Data}
\label{sub:pipeline}
Detector-level processing of JWST H2RG exposures is done in the first pipeline
stage, {\tt calwebb\_detector1}, which processes the data from non-destructively
read integration ramps to slope images (with count rate units of ADU/s). The
first steps flag the dead, hot, noisy, and saturated pixels, followed by the
subtraction of a superbias frame and a reference pixel correction which corrects
for drifts between rows and columns of the charge injected by the readout
electronics. A non-linearity correction and optional persistence correction are
then applied, followed by subtraction of the dark signal. Jumps (such as those
caused by cosmic ray hits) are then flagged in the so-called {\tt jump} step
which uses the two-point difference method described in \citet{andgor11}. The
final step in {\tt calwebb\_detector1} is the {\tt ramp\_fitting} step which fits
a slope to the reads of each pixel during an integration ramp, after discarding the
reads that were flagged during the previous steps. If an exposure contains
multiple ramps, a file with the slope averaged over all integrations is
also produced.

During the second pipeline stage, called {\tt calwebb\_image2}, world coordinate
system (WCS) and flux calibration information is added to the file and a
flat-field correction is applied. Finally, {\tt calwebb\_image2} resamples the
input image into a distortion-free product, using the WCS and distortion
information added earlier. By default, the input-to-output pixel mapping applied
during this {\tt resample} step uses the IVM weighting scheme that uses the
inverse of the read noise variance array that was stored in each image during the
{\tt ramp\_fitting} step in {\tt calwebb\_detector1}. Relevant suffix names of
output files of the {\tt calwebb\_detector1} and {\tt calwebb\_image2} pipelines
are listed in Table~\ref{tab:pipelinefiles}. 

The third and last pipeline stage, {\tt calwebb\_image3}, combines multiple
exposures (e.g., all dither positions in a dithered exposure sequence) taken
with a given filter into a single drizzled image, using the same resampling and
weighting scheme as that mentioned above during the {\tt calwebb\_image2} stage.

The impact of BFE on science with point source imaging data of undersampled PSFs
with H2RG detectors is mainly due to the logic of the {\tt jump} and {\tt
  ramp\_fitting} steps of the {\tt calwebb\_detector1} pipeline stage. This will
be described in detail in the next Section. 

\begin{table*}[htb]
 \caption{JWST Calibration Pipeline Output File Nomenclature \label{tab:pipelinefiles}}
 \footnotesize
 \vspace*{-2ex}
\begin{center}
 \begin{tabular}{@{}lrp{9cm}@{}} \tableline \tableline
 Pipeline Name & File Suffix & Notes  \\ [0.5ex] \tableline
 {\tt calwebb\_detector1} & {\tt \_ramp.fits} & Calibrated ramps just before
                                                ramp slope fitting. Only produced when
                                                setting {\tt
                                                  save\_calibrated\_ramp = True}. \\
  & {\tt \_rateints.fits} & Calibrated count rate (one per ramp) \\ 
  & {\tt \_rate.fits} & Calibrated count rate (averaged over all ramps) \\ 
  & &  \\ [0.2ex]  
 {\tt calwebb\_image2} & {\tt \_cal.fits} & WCS added and flat-field correction
 applied; pixel values converted to surface brightness in MJy/sr \\
  & {\tt \_i2d.fits} & Resampled onto distortion-free image array. Only produced
 when setting {\tt resample.single = True} and {\tt resample.skip = False}. \\ [0.5ex] \tableline
\end{tabular}
\end{center}
\end{table*}

\section{Impact of BFE to NIRISS H2RG Images}
\label{s:impact}

\subsection{Undersampled PSFs}
\label{sub:undersamp}

\begin{figure*}[p]
  \centerline{\includegraphics[width=16cm]{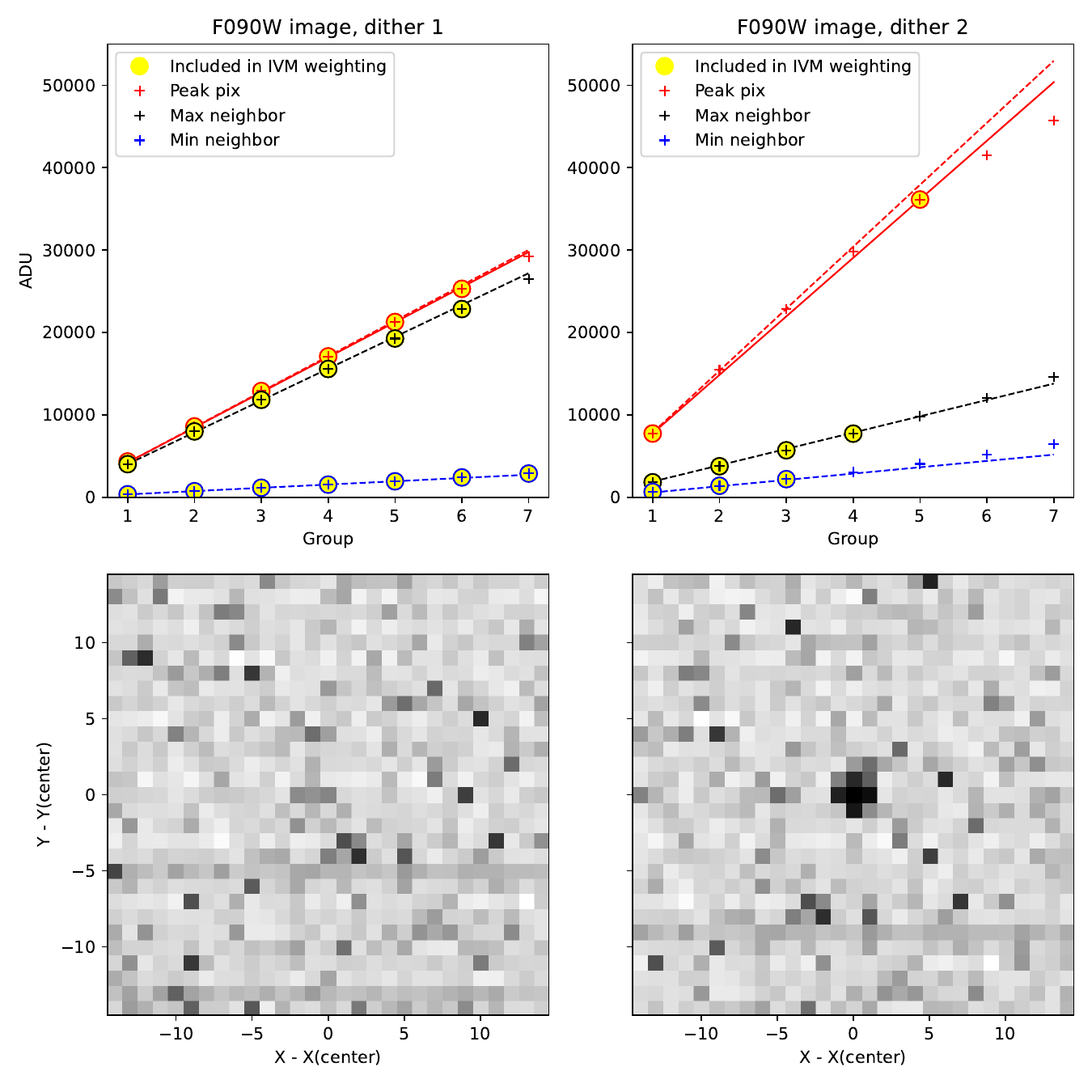}}
  \caption{Illustration of impact of BFE for well-illuminated stars with
    spatially undersampled PSF in an imaging exposure sequence using the F090W
    filter at two dither positions. Left and right panels are for dither
    positions 1 and 2, respectively. 
    {\it Top panels:} linearized pixel ramps (plus signs) for the peak pixel
    (red) and the neighboring pixels with the maximum and minimum signal levels
    (black and blue, respectively). Dashed lines show linear fits to the groups
    for which the peak pixel accumulated linearized counts $<$ 25,000 ADU.
    For comparison, the red solid line represents the ramp slope for the peak
    pixel calculated by {\tt calwebb\_detector1} with default parameter
    settings, and the open circles indicate groups that did {\it not} get
    flagged during the {\tt jump} step. Yellow circles indicate groups
    that are included in the IVM weighting process in the {\tt
      resample} step. 
    Note the charge migrating from pixels with accumulated signal level $\ga$
    25,000 ADU to neighboring pixels with lower intrinsic signal levels.
    {\it Bottom panels:} greyscale weight maps as derived from the
    {\tt VAR\_RNOISE} extension of the {\tt \_rate} files, used by the IVM
    weighting scheme to resample images into distortion-free products by the
    drizzle algorithm. Darker color indicate lower IVM weights.  
    Note the low weights assigned to the central pixels of the star in dither
    position 2. See Section~\ref{sub:undersamp} for discussion. 
    \label{fig:BFE_F090W}
  }
\end{figure*}

A good example of the significant impact BFE can have on point source science
with undersampled PSFs is provided by the dataset for exposure specification
\#\,9 of JWST program 1094 (PI: A. Martel). This dataset consists of F090W images
of a flux standard star (LDS\,749, a DBQ4 white dwarf, cf.\ \citealt{bohkoe08})
taken at two dither positions that differ in pixel phase $\phi$ by
($\Delta\phi_{\rm x}$, $\Delta\phi_{\rm y}$) = (0.5, 0.5) pixels.
In this particular case, the star was centered near a pixel corner in the
first dither position and near a pixel center in the second position. As shown
in Figure~\ref{fig:BFE_F090W}, this setup resulted in the peak pixel reaching
a count rate $\sim$\,50\% higher in dither position 2 than in position
1. Obviously, the F090W PSF is strongly undersampled by the NIRISS detector.

The plus signs in the top panels of Figure~\ref{fig:BFE_F090W} show the
linearized count levels attained during the integration ramp of those two
images, for their peak pixels and two neighboring pixels. For comparison, two
lines are drawn for the peak pixels: the solid line depicts the ramp slope
calculated by the {\tt ramp\_fitting} step in {\tt calwebb\_detector1}, while
the dashed line depicts a linear fit to the first three reads (hereafter
referred to as ``groups'' following the JWST nomenclature) of the ramp. Note
that the two slopes are virtually identical for the first dither position and
fit the data very well (and this is also the case for the neighboring
pixels), while the signal levels of the data for the second dither position get
progressively below the dashed line at accumulated signal levels $\ga\,$25,000
ADU. This is the BFE, and it is accompanied by signal levels in the
neighboring pixels that are \emph{above} their respective linear fits to the
groups of the ramp for which the peak pixel stays below $\sim$\,25,000 ADU. This
``surplus charge'' in the pixels next to the peak pixel represents charge that
migrated from the peak pixel to its neighbors with significantly lower signal
levels.

\begin{figure}[htb]
  \centerline{\includegraphics[width=8.8cm]{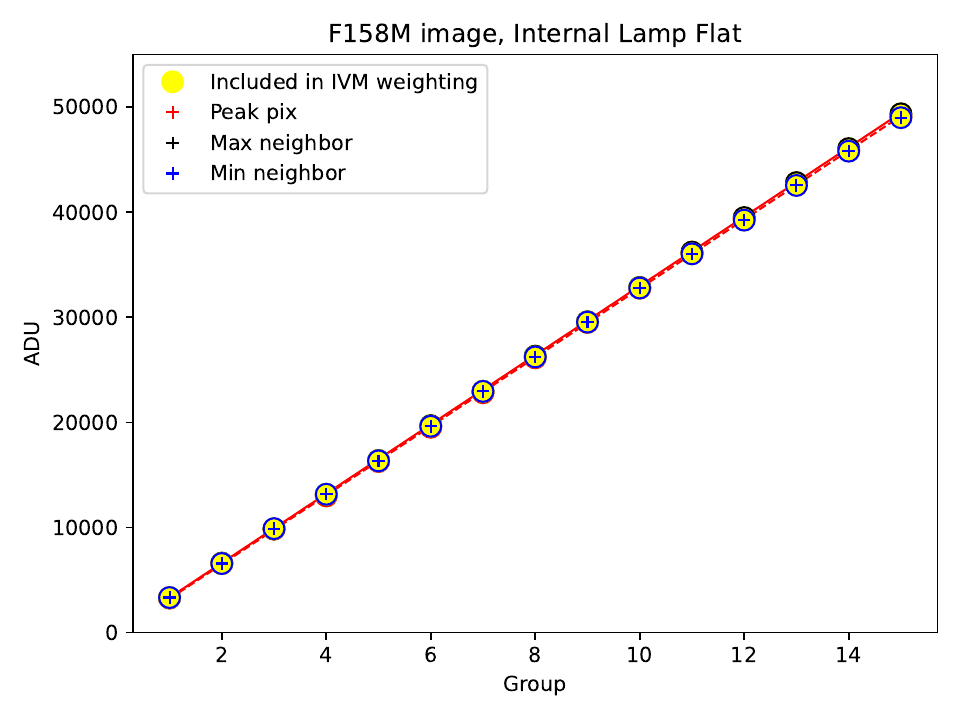}}
  \caption{Same as top panels in Figure~\ref{fig:BFE_F090W}, but now for
    exposure jw01083001001\_0210a\_00001\_nis\_uncal.fits, an image illuminated
    by a lamp internal to NIRISS, for 3 adjacent pixels in an uniformly
    illuminated part of the image. Note the absence of any appreciable BFE effect
    on such data. The count rate levels of the 3 adjacent pixels are the same to
    within $\sim$\,1\%. 
    \label{fig:flatslopes}
  }
\end{figure}

Before going into details regarding the impact of BFE to science with
undersampled images, it is important to realize that the linearity correction
that is applied to JWST H2RG data in the calibration pipeline is \emph{not}
affected by the BFE. This linearity correction 
is derived from a set of images taken with a dedicated external lamp during
ground testing, providing stable and uniform illumination of the flight 
detector  \citep[see][]{morishita+20}. The BFE has no detectable effect on such
uniformly illuminated data, as illustrated in Figure~\ref{fig:flatslopes}.

\begin{figure*}[htb]
  \centerline{\includegraphics[width=13.5cm]{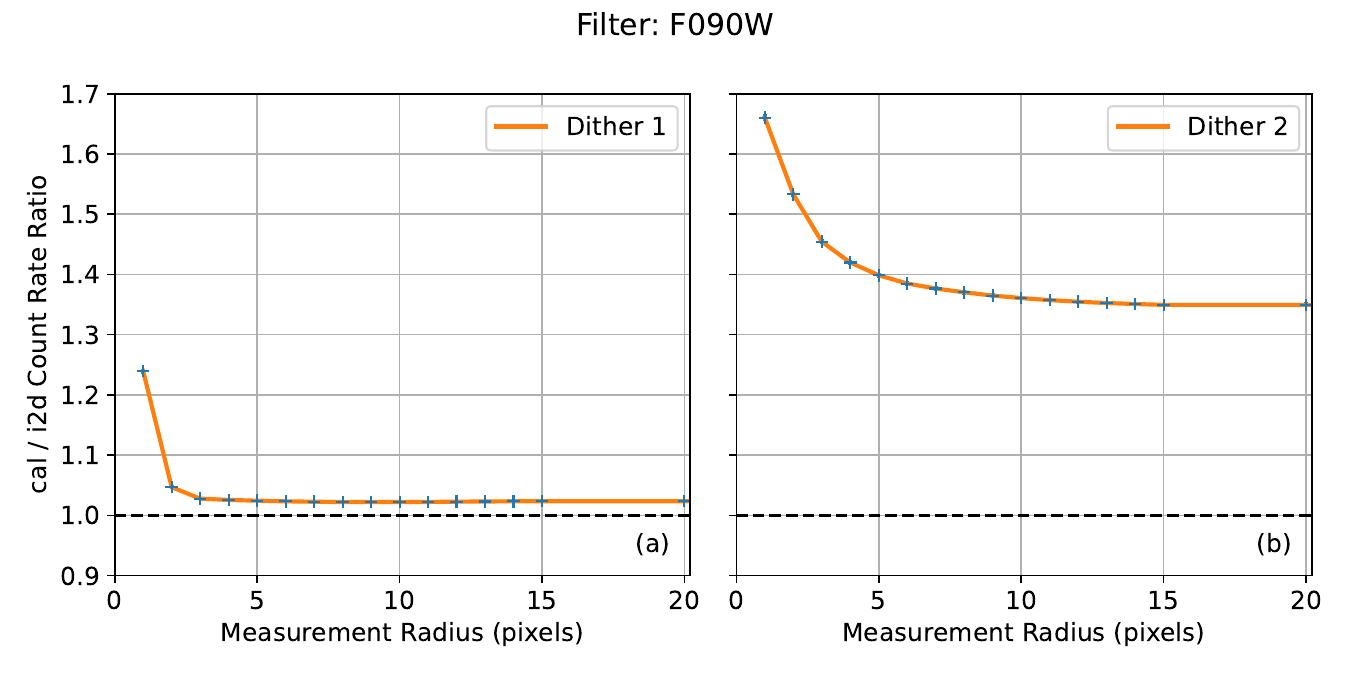}}
  \caption{``cal\,/\,i2d'' ratio of integrated count rates measured from the
    {\tt \_cal} and {\tt \_i2d} pipeline products as a function of circular
    aperture radius for the dataset shown in Figure~\ref{fig:BFE_F090W}. 
    Panels (a) and (b) are for dither positions 1 and 2, respectively. Plus signs
    represent aperture photometry measurements which are connected by the solid
    lines. 
    Note that the resampling process for the star in dither position 1
    (for which only one group in the central pixels suffered from the
    BFE) only changed its integrated count rate by a small
    amount ($\sim$\,2\%), as  expected. However, the resampling
    process for the same star in dither position 2 resulted in a net
    loss of integrated count rate of $\sim$\,36\% as a consequence of
    the low IVM weights assigned to the central few pixels of the PSF
    (see bottom right panel of Figure~\ref{fig:BFE_F090W}) due to the
    significant BFE. 
    \label{fig:cal_i2d_F090W}
  }
\end{figure*}

The impact of BFE to the derived ramp slopes for stars in undersampled H2RG data is
mainly due to the flagging done in the {\tt jump} step within {\tt 
  calwebb\_detector1}, which iteratively flags groups $N$ if  
the absolute two-point difference $\left|{\it group}_N - {\it group}_{N-1}\right|$
is larger than the median absolute two-point difference of the full ramp by a
certain threshold; this threshold is defaulted at $4\,\sigma$ where $\sigma$
refers to the read noise for two groups. This default threshold value was found
from testing to provide solid flagging of cosmic ray hits
\citep[see][]{andgor11}.  
However, the {\tt jump} step introduces negative side effects for exposures that
suffer from significant BFE. Taking the second dither position of the dataset
presented here as an example, the {\tt jump} step assigned flags \emph{to all
groups except \#\,1 and 5} in case of the peak pixel, while jump flags were
assigned to groups $\geq$\,4 in the neighboring pixels. This
caused two problems: 
\\ [-3.2ex] 
\begin{enumerate}
  \item The ramp slope calculated by the pipeline for the peak pixel is lower
    (in this case by 4.5\%) than that calculated from the groups with
    accumulated signal levels of $\la$\,25,000 ADU, while the charge that
    migrated from the peak  pixel to the surrounding pixels \emph{is not used 
    by the ramp slope calculations} for the latter pixels, since that surplus
    charge is flagged as jumps, and groups with jumps are excluded from ramp
    slope calculations. As a result, the integrated flux for the star is skewed
    low, which was noticed during NIRISS commissioning when comparing the
    measured integrated fluxes from the two dithered exposures.
  \item Perhaps more importantly for science applications that involve image
    combination of distortion-free images using the drizzle algorithm, the
    flagging of multiple groups by the {\tt jump} step in pixels affected by BFE
    can cause \emph{significant loss of flux} in the resampled and combined
    images. This is due to the IVM weighting used to resample images onto a
    distortion-free pixel grid within the drizzle algorithm to combine
    dithered images in the {\tt calwebb\_image3} pipeline.   
    IVM weights are derived for each pixel as $({\it var}_{\it RNOISE})^{-1}$
    where ${\it var}_{\it RNOISE}$ is the variance of the slope of a ramp (or
    ramp segment) due to read noise (see
    \href{https://tinyurl.com/23r86jzz}{ReadTheDocs article for the {\tt
        ramp\_fitting} step} for details), which is represented by the {\tt
      VAR\_RNOISE} extension of the {\tt \_rate} and {\tt\_cal} pipeline
    products. 
    The IVM weight maps of the 30\,$\times$\,30 pixel region around the star in
    the two dither positions in the dataset discussed here are shown in the
    bottom panels of Figure~\ref{fig:BFE_F090W}. Note the low IVM weights
    assigned to the central pixels of the star in the second dither position
    (i.e., the one significantly affected by BFE), which are due to significant 
    numbers of groups getting flagged as ``jumps'' in that image for those pixels. 
    The consequence of these low IVM weights for the pixels with high signal
    level is that the {\tt resample} step, which resamples the input image onto
    a distortion-free pixel grid, effectively lowers the pixel values in the
    PSF region in the output {\tt \_i2d} image relative to the situation in the
    input image. This is readily seen when comparing aperture photometry of
    the star from the {\tt \_cal} and {\tt \_i2d} pipeline products using
    multiple measurement radii: Figure~\ref{fig:cal_i2d_F090W} shows
    that the integrated count rate of the star in the {\tt \_i2d} image of the
    second dither position is lower by $\sim$\,36\% relative to the input {\tt
      \_cal} image, while the {\tt \_cal} and {\tt \_i2d} count rates of the
    star in the first dither position (for which only one group in the central
    pixels was affected by BFE) are consistent with each other to within
    2\%. \\ [-3.5ex] 
\end{enumerate}

\subsection{Adequately Sampled PSFs}
\label{sub:wellsamp}
For purposes of comparison with the case of undersampled PSFs, we now
illustrate the impact of BFE on images with adequately sampled PSFs, using
filter F480M for which the PSF has a FWHM of $\sim$\,2.5 pixels. This is
done using direct images taken during observation \#\,23 of JWST
program 1093 (PI: D. Thatte). This dataset consists of F480M images of
CPD$-$67$-$607, a bright K giant star used as PSF reference star,
taken in two dither positions, again differing in pixel phase by
($\Delta\phi_{\rm x}$, $\Delta\phi_{\rm y}$) = (0.5, 0.5) pixels. In this case
the star was centered very close to a pixel center in the first dither position
and near a pixel corner in the second position. The difference of the measured
count rates for the peak pixel between the two dithers is only 
$\sim$\,14\% in this case, as opposed to $\sim$\,50\% for 
the strongly undersampled F090W images described in Sect.\ \ref{sub:undersamp}.

\begin{figure*}[htb]
  \centerline{\includegraphics[width=16cm]{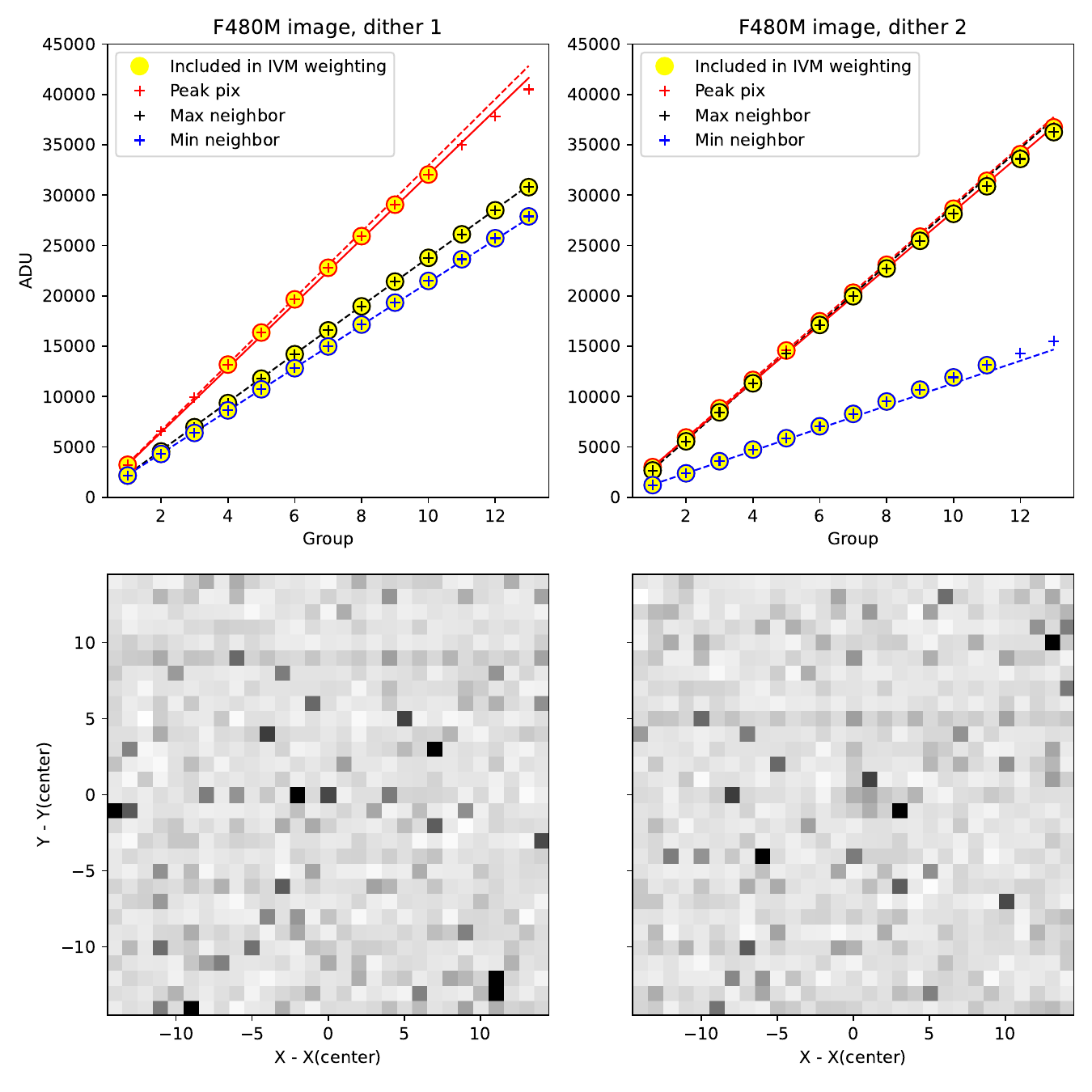}}
  \caption{Same as Figure~\ref{fig:BFE_F090W}, but now for a spatially
    adequately sampled star in an imaging exposure sequence using the F480M
    filter at two dither positions. See Section~\ref{sub:wellsamp} for discussion. 
    \label{fig:BFE_F480M}
  }
\end{figure*}

The linearized ramps and IVM weight maps for the two F480M images as determined
by the operational pipeline are shown in Figure~\ref{fig:BFE_F480M} which has the
same setup as Figure~\ref{fig:BFE_F090W}. For the F480M exposure in which the
PSF was centered on a pixel, the BFE caused the {\tt jump} step to flag 5 out of
the 13 groups up the ramp for the peak pixel, yielding a ramp slope that is lower
by 2.1\% than the slope calculated from the groups with accumulated signal
levels of $\la$\,25,000 ADU (see top left panel of Figure~\ref{fig:BFE_F480M}).
This is reflected in a relatively low IVM weight for the peak pixel (see bottom
left panel). However, in contrast with the case of the undersampled PSF in the
F090W image that was centered on a pixel, the amount of charge that was migrated
to the adjacent pixels due to the BFE was too low to cause jump detections
there, likely because of the relatively low contrast in signal level between the
peak pixel and its neighbors. For the F480M exposure that was centered 
near a pixel corner (see right-hand panels of Figure~\ref{fig:BFE_F480M}), the
peak pixel and one adjacent pixel had very similar 
count rates and only one of the 13 groups was flagged by the {\tt jump} step,
while the neighboring pixel with the lowest count rate did receive a detectable
amount of migrated charge (as evidenced by jump detections) in the last two
groups, likely because of the relatively large contrast in signal level with its
neighbors.  As such, several pixels near the center of the PSF received a small
but non-negligible lowering of IVM weights (see bottom left panel of
Figure~\ref{fig:BFE_F480M}). Furthermore, the ramp slope for the peak pixel 
calculated by the operational pipeline for this exposure is 1.8\% lower than
that calculated from the groups with accumulated signal levels of $\la$\,25,000
ADU. 

\begin{figure*}[htb]
  \centerline{\includegraphics[width=13.5cm]{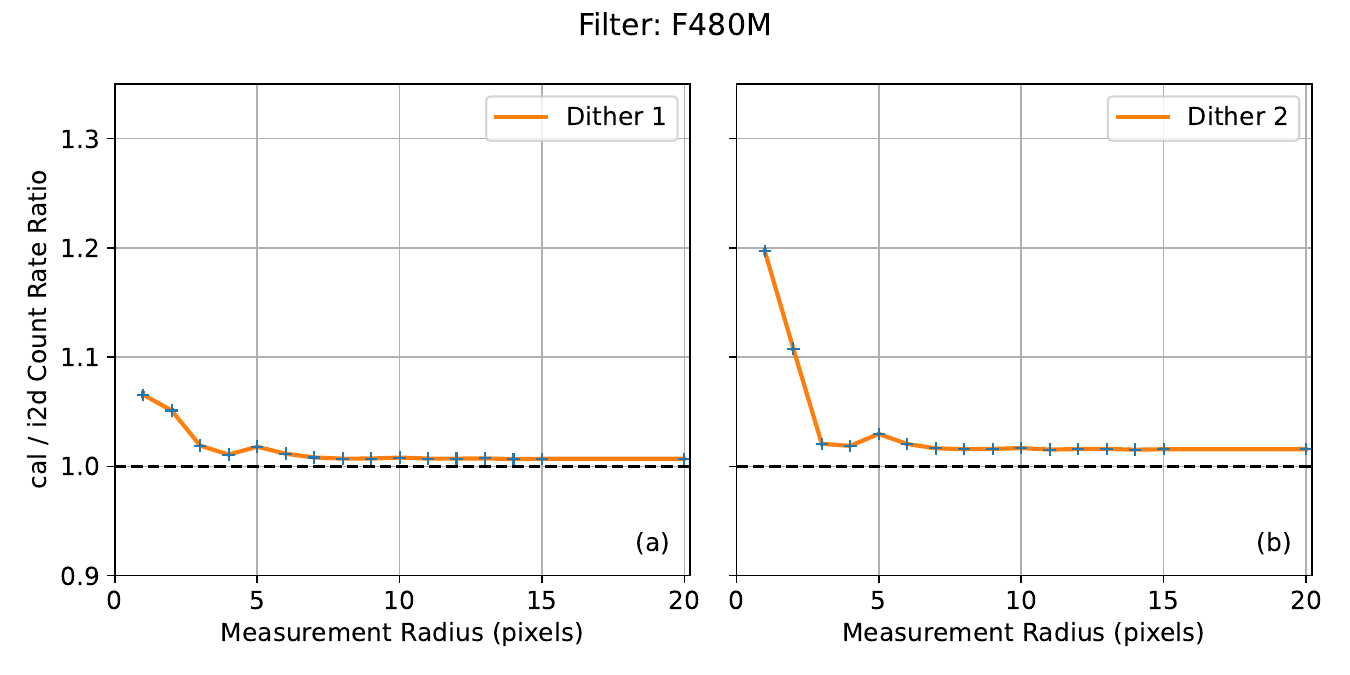}}
  \caption{Same as Figure~\ref{fig:cal_i2d_F090W}, but now for the F480M dataset
    shown in Figure~\ref{fig:BFE_F480M}.  
    The resampling process in the {\tt resample} step for this star resulted in a
    net loss of integrated count rate of $\sim$\,2\% in both dither positions due
    to the lowered IVM weights assigned to the central 1-5 pixels of
    the PSF (see bottom panels of Figure~\ref{fig:BFE_F480M}).
    \label{fig:cal_i2d_F480M}
  }
\end{figure*}

The impact of the IVM weights for these F480M images is shown in terms of the
``{\tt cal/i2d}'' ratio of integrated count rates in
Figure~\ref{fig:cal_i2d_F480M}. The resampling process used to create the {\tt
  \_i2d} image resulted in a net loss of integrated count rate of $\sim$\,2\%
for both dithers, due to the jump detections in some groups for the pixels at or
near the PSF centers as described above. While this count rate loss for
adequately sampled PSFs due to the BFE is significantly lower than it is in
undersampled data with similar levels of total charge, it is still significant
and relevant to mitigate the issue to improve photometric precision. 

\section{The \texttt{charge\_migration} Algorithm}
\label{s:algorithm}

\subsection{Description}
\label{sub:description}

To address the issues caused by the BFE for undersampled H2RG data described in
the previous section, we designed an additional step within the {\tt
  calwebb\_detector1} pipeline called {\tt charge\_migration} which is
inserted in between the {\tt dark\_current} and {\tt jump}
steps.  The new step has one input parameter {\tt signal\_threshold}, which has
a default value of 25,000 ADU which can be replaced by other values for a given
exposure type or set of optical elements by means of
parameter reference files. The determination of the values of {\tt
  signal\_threshold} for NIRISS modes is described in the Appendix.

The {\tt charge\_migration} step assigns a data quality (DQ) flag called {\tt
  CHARGELOSS} to any non-saturated group in any integration whose accumulated
signal is above the value of {\tt  signal\_threshold}. Furthermore, the same DQ
flag is also assigned to the same groups of the pixels that are immediate
neighbors of those high-signal pixels. This is done to ensure that the
inclusion or exclusion of groups from calculations in the {\tt jump} and {\tt
  ramp\_fitting} steps is done in the same way for pixels with values above
{\tt signal\_threshold} and their neighbors who receive ``surplus'' charge
that is migrated from the high-signal pixel due to the BFE. 
The presence of the {\tt CHARGELOSS} DQ flag (which are saved in the {\tt
  GROUPDQ} array associated with the data file) results in certain actions in
the subsequent {\tt jump} and {\tt ramp\_fitting} steps, both of which have been
updated along with the implementation of the {\tt charge\_migration}
step\footnote{As of version 1.12.3 of the {\it jwst} python package and CRDS
context 1135, the {\tt charge\_migration} step has been activated for data taken
with NIRISS observing modes AMI, Imaging, and WFSS.}. These actions are as follows:  
\begin{itemize}
\item in the {\tt jump} step, the groups with the {\tt CHARGELOSS}
    DQ flag are being excluded from the jump detection calculations (i.e., the
    two-point difference calculations), similar to groups with the {\tt
      SATURATION} DQ flag. The groups with the {\tt CHARGELOSS} flag are
    therefore \emph{not} issued a {\tt jump} DQ flag.
\item in the subsequent {\tt ramp\_fitting} step, the groups with the {\tt
  CHARGELOSS} flag are being excluded from the ramp slope calculations. However,
  since those groups were not assigned a {\tt jump} DQ flag, they \emph{are} 
  included in the calculation of the variance of the slope due to read
  noise (i.e., the {\tt VAR\_RNOISE} array). This prevents them from being
  assigned a low IVM weight during the {\tt resample} step in the {\tt
    calwebb\_image2} and {\tt calwebb\_image3} pipelines, which they were
  before the implementation of the {\tt charge\_migration} step.
\end{itemize}

\begin{figure*}[htb]
  \centerline{\includegraphics[width=16cm]{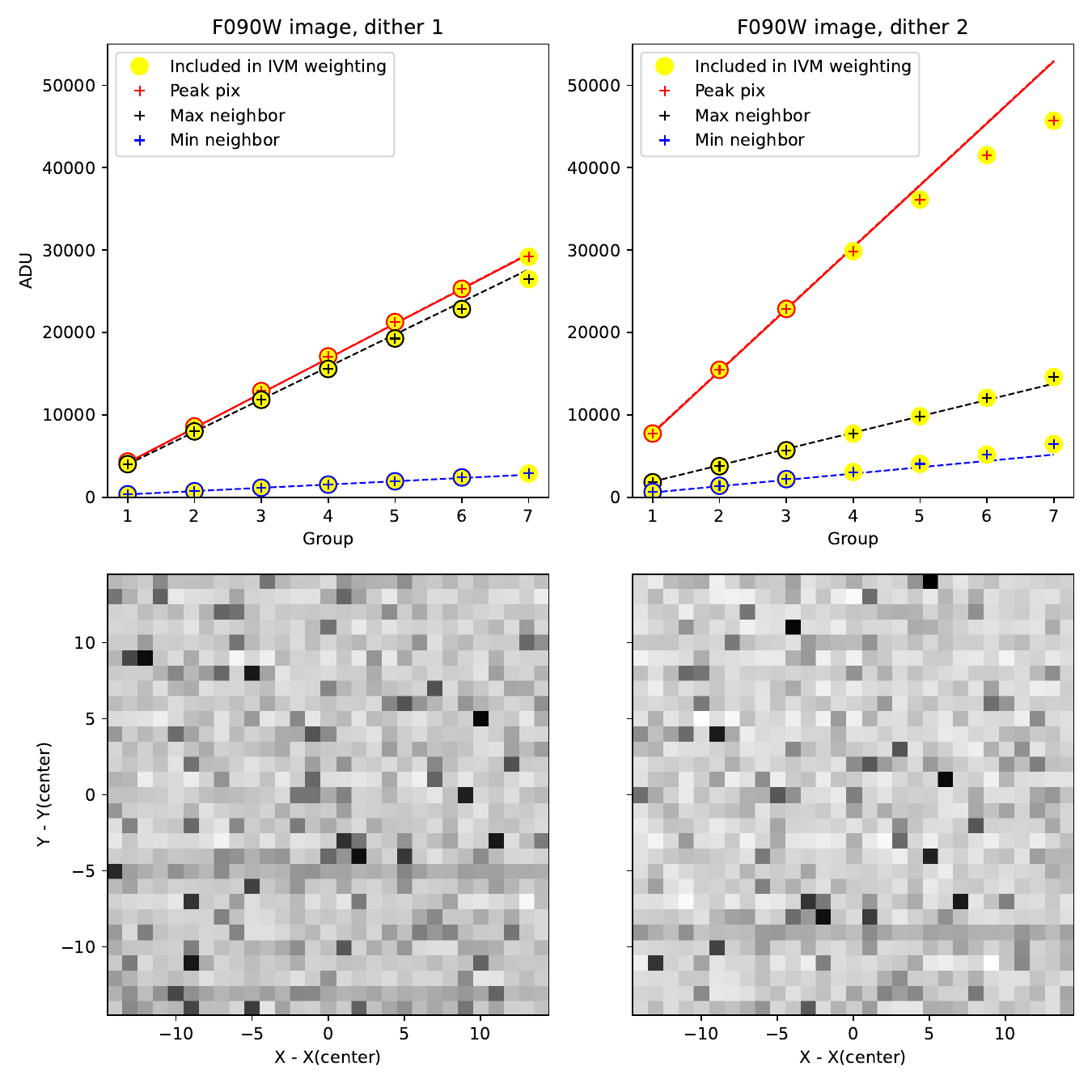}}
  \caption{Same as Figure~\ref{fig:BFE_F090W}, but now
    illustrating the situation after the {\tt charge\_migration} step
    was implemented and activated. In the top panels, note that the {\tt jump}
    step now correctly avoids flagging the groups for which the peak pixel
    accumulated linearized counts $<$ 25,000 ADU (if they do not show actual
    jumps, that is). This is especially relevant for dither position 2 (see
    right-hand panel).  Note also that all (non-saturated) groups are now
    included in the IVM weighting process in the {\tt resample} step. 
    In the bottom panel, note that the low IVM weights that were assigned to the 
    central pixels of the star in dither position 2 prior to the implementation
    of the {\tt charge\_migration} step have now disappeared. See
    Section~\ref{s:algorithm} for discussion. 
    \label{fig:BFE_F090W_after}
  }
\end{figure*}

We remind the reader that the main purpose of the use of IVM weights based on
read noise during image combination (as opposed to weighting by exposure time) is
to optimize sensitivity for faint objects \citep[see][]{caser+00}. The
modification of the IVM weighting scheme performed by the {\tt
  charge\_migration} step only affects the brightest objects in images, for
which the read noise is typically negligible relative to the poisson
noise associated with the source\footnote{In our tests, we find
${\it var}_{\it RNOISE}$/{\it var}$_{\it POISSON} < 0.05$ for all flagged
pixels.}. As such, the benefits of IVM weighting are fully retained when
running the {\tt charge\_migration} step.  

\subsection{Results}
\label{sub:results}

The impact of the {\tt charge\_migration} step in addressing the effect of the
BFE on science for H2RG data can be significant, especially in the case
of spatially undersampled data of bright stars. This is illustrated in this
section using files produced by re-running the {\tt calwebb\_detector1}
pipeline on the {\tt \_uncal.fits} files, now with the {\tt charge\_migration}
step activated.

\subsubsection{Undersampled data}
\begin{figure*}[htb]
  \centerline{\includegraphics[width=13.5cm]{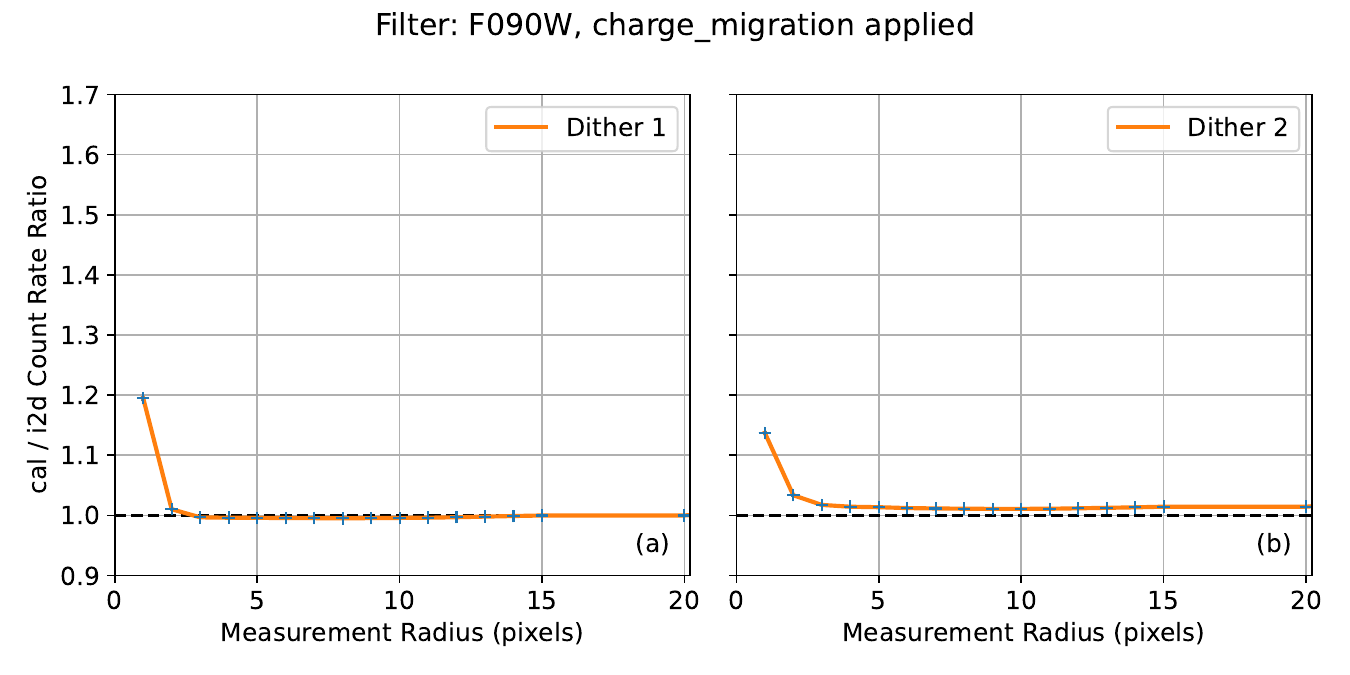}}
  \caption{Same as Figure~\ref{fig:cal_i2d_F090W}, but now illustrating
    the situation after the {\tt charge\_migration} step was implemented
    and activated. Scale of the ordinate is the same as in
    Figure~\ref{fig:cal_i2d_F090W}. The integrated count rate of the star
    in dither position 2 after the {\tt resample} step is now the same as that
    from the flat-fielded ({\tt \_cal}) image to within 1\%\ as opposed to being
    $\sim$\,36\% lower without the {\tt charge\_migration} step
    (cf.\ Figure~\ref{fig:cal_i2d_F090W}).  
    \label{fig:cal_i2d_F090W_after}
  }
\end{figure*}

Figure~\ref{fig:BFE_F090W_after} is the same as Figure~\ref{fig:BFE_F090W} for
the F090W dataset, but now showing the ramps, ramp-fitting results, and IVM
weight assignments of the pixels around the star in the two dither positions in
the data produced after activating the {\tt charge\_migration} step. Note that
the inappropriate assignments of low IVM weights to the central pixels of the
PSF have now disappeared. Furthermore, the integrated flux measurements of the
star in the resampled {\tt \_i2d} images are now consistent with those in the
flat-fielded {\tt \_cal} images to within 1\% for both dither positions, as
shown in Figure~\ref{fig:cal_i2d_F090W_after}.   

\begin{figure*}[ptb]
  \centerline{\includegraphics[width=16cm]{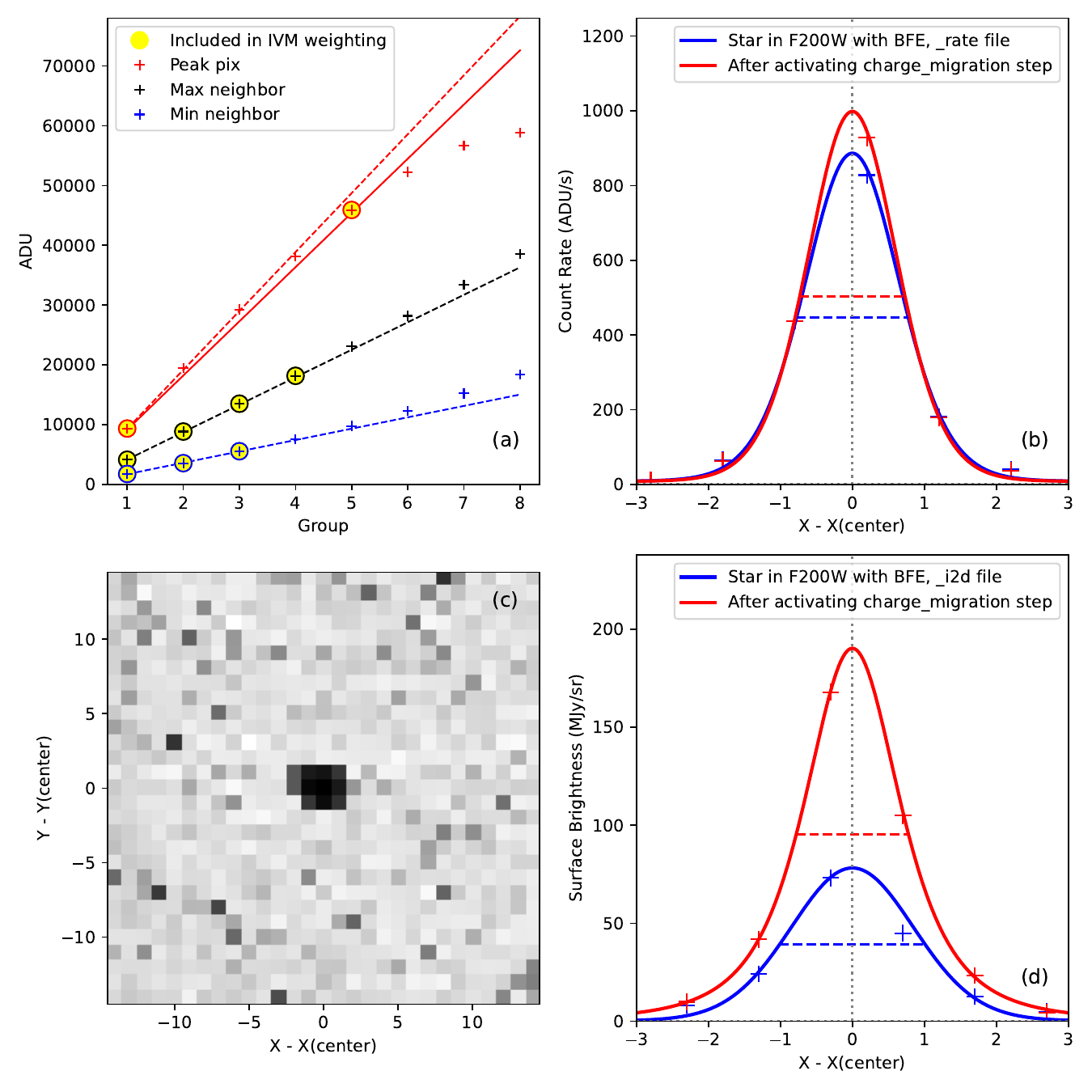}}
  \caption{Impact of the new {\tt charge\_migration} step on the BFE
    effect on a star in NIRISS exposure {\tt jw01096001001\_02101\_00001}, a
    F200W imaging exposure with NGROUPS = 8 described in Section~\ref{s:algorithm}.
    {\it Panel (a)}: same as top panels in Figure~\ref{fig:BFE_F090W} but
    now for this exposure. {\it Panel (b)}: the blue and red plus signs show the X
    crosscut of the PSF profile of this star in {\tt \_rate} images without
    and with the {\tt charge\_migration} applied, respectively. The blue
    and red solid lines represent best-fit EFF functions \citep{eff87} to those
    data. FWHM values are indicated. The {\tt charge\_migration} step
    increased the flux of the peak pixel in the {\tt \_rate} image by 13\%, and
    the integrated flux by 2\%. 
    {\it Panel (c)}: same as bottom panels in Figure~\ref{fig:BFE_F090W} but
    now for this exposure. Note the low IVM weights for the inner $3\times3$
    pixels of the PSF due to the BFE effect.
    {\it Panel (d)}: Same as panel (b), but now for the {\tt \_i2d} images. In
    this case, the {\tt charge\_migration} step increased the flux of
    the peak pixel by 144\% and the integrated flux of the star by 52\%. 
    \label{fig:BFEplot_f200w}
  }
\end{figure*}

Another effect of BFE 
is that it causes an apparent widening of the PSF due to the peak signal being
depressed relative to that of the surrounding pixels. As already described in
Section~\ref{s:impact} and shown in Figure~\ref{fig:BFE_F090W}, the
efficacy of the BFE scales with pixel-to-pixel contrast, which for
undersampled PSFs includes the placement of PSFs within pixel
boundaries. As such, this issue affects efforts to create proper sets of
empirical PSFs \citep[ePSFs; see, e.g.,][]{andkin00,andkin06,libralato+23} for
high-precision astrometry and photometry of point sources in undersampled H2RG
data from JWST.  The effect on PSF width and shape is again most significant in
the resampled {\tt \_i2d} files due to the IVM weighting described above. To
illustrate this effect, we use an F200W image with a ramp of 8 groups taken as
part of JWST Commissioning Program 1096 (PI: A. Martel) in which a star is
centered very close to a pixel center. The data for this star show a particularly
strong efficacy of the BFE because the signal level of the peak  pixel almost
reaches the saturation threshold at the last group, causing a particularly
strong downward curvature of the (linearity-corrected) ramp beyond  the {\tt
  signal\_threshold} value, which is already reached at the third group. As
shown in panel (a) of Figure~\ref{fig:BFEplot_f200w}, the {\tt jump} step causes
the ramp slope to be derived from only 2 groups for the peak pixel (one of which
is significantly affected by BFE), and from 3 (different) groups for its
neighbors. This in turn causes very low IVM weight assignments for the central
pixels as shown in panel (c) of Figure~\ref{fig:BFEplot_f200w}.   

The effect of this to the PSF shape in the {\tt \_cal} and {\tt \_i2d} images is
shown in panels (b) and (d) of Figure~\ref{fig:BFEplot_f200w}, respectively,
both before and after the {\tt charge\_migration} step was implemented
into the pipeline. In the case of the flatfielded {\tt \_cal} image, the
application of the {\tt charge\_migration} step yielded a flux increase
of 13\% in the peak pixel, the integrated flux of the star was increased by 2\%,
and the measured FWHM of the PSF showed a moderate decrease (from 1.55 to 1.46
pixels). However, for the resampled {\tt \_i2d} images, the differences are
quite dramatic: the peak pixel flux increased by 144\%, the integrated flux 
increased by 52\%, and the FWHM decreased significantly as well (from
2.02 to 1.56 pixels).  As such, the implementation of the {\tt
  charge\_migration} step yields a significant improvement of the quality and
internal consistency of PSFs and PSF (or ePSF) reference libraries for
PSF-fitting photometry of undersampled JWST imaging modes such as NIRISS imaging
at wavelengths $\la 2\;\mu$m and NIRISS AMI data.    

\subsubsection{Adequately Sampled Data}

\begin{figure*}[htb]
  \centerline{\includegraphics[width=16cm]{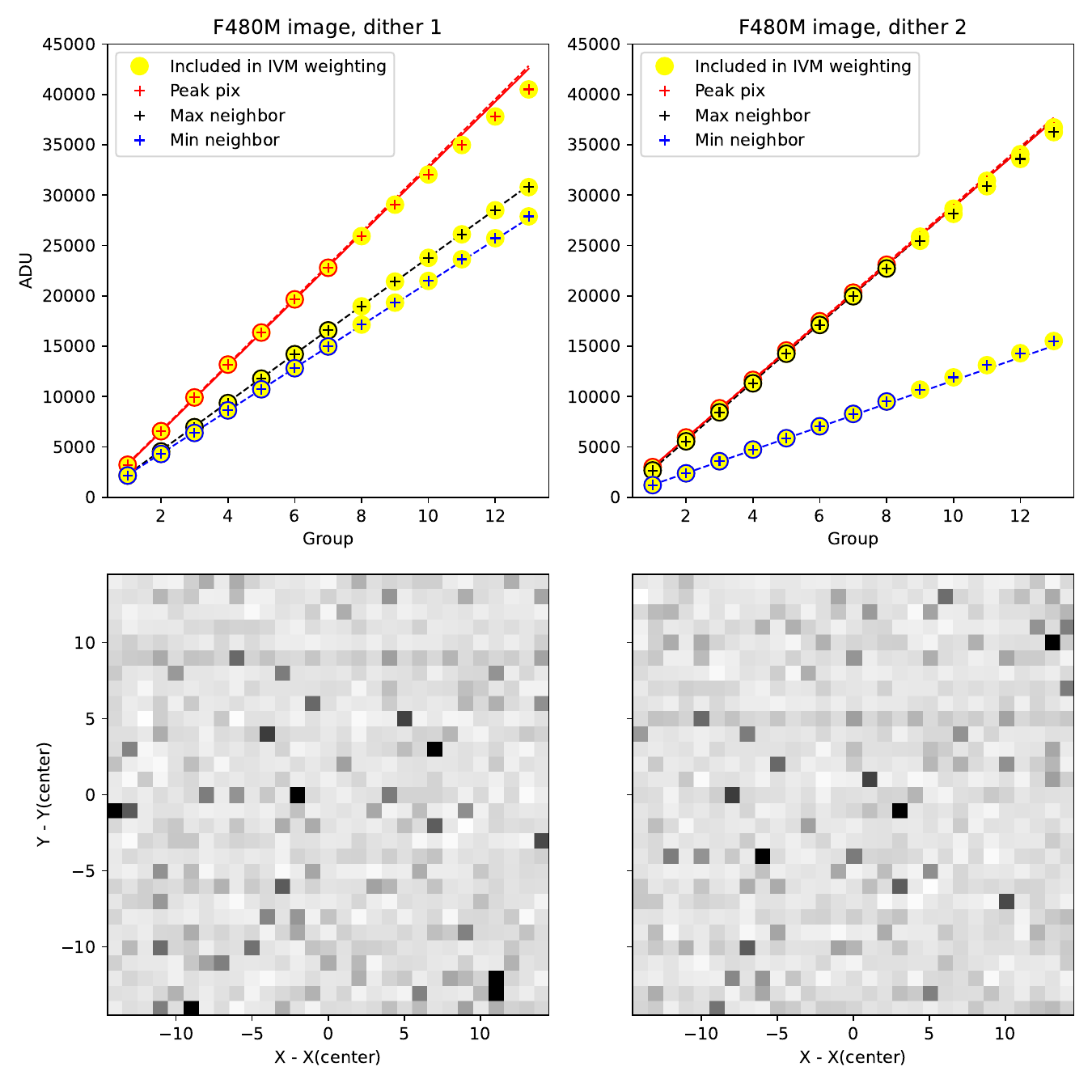}}
  \caption{Same as Figure~\ref{fig:BFE_F480M}, but now
    illustrating the situation after the {\tt charge\_migration} step
    was implemented and activated. Note that no jump detections remain
    in the central pixels of the PSFs, and hence all groups are now
    included in the IVM weighting process in the {\tt resample} step. 
    See Section~\ref{sub:results}.2 for discussion. 
    \label{fig:BFE_F480M_after}
  }
\end{figure*}

\begin{figure*}[htb]
  \centerline{\includegraphics[width=13.5cm]{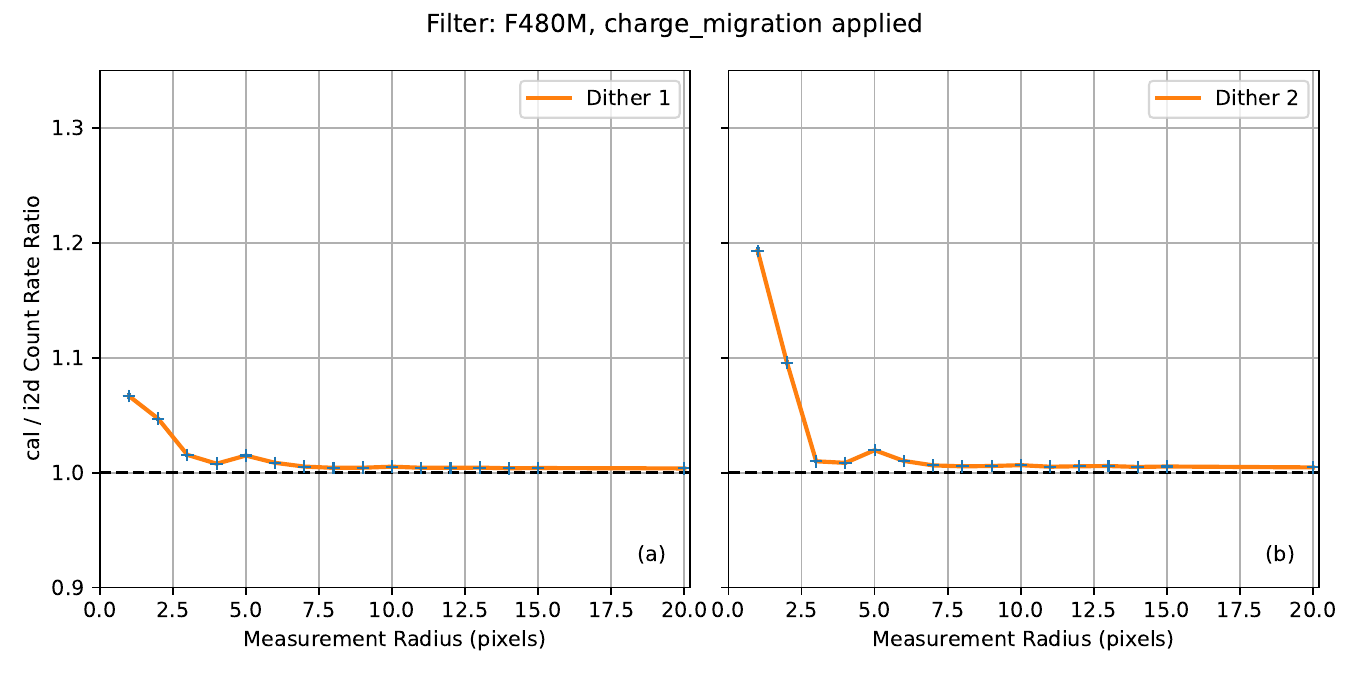}}
  \caption{Same as Figure~\ref{fig:cal_i2d_F480M}, but now illustrating
    the situation after the {\tt charge\_migration} step was implemented
    and activated. Scale of the ordinate is the same as in
    Figure~\ref{fig:cal_i2d_F480M}. The integrated count rate of the star
    after the {\tt resample} step is now the same as that
    from the flat-fielded ({\tt \_cal}) image to within 0.5\% (for
    both dither positions) as opposed to being
    $\sim$\,2\% lower when processed without the {\tt charge\_migration} step.  
    \label{fig:cal_i2d_F480M_after}
  }
\end{figure*}

The ramps, ramp-fitting results, and IVM weight assignments of the pixels around
the star in the two dither positions in the F480M images produced after activating
the {\tt charge\_migration} step are shown in
Figure~\ref{fig:BFE_F480M_after}. Similar to the case of the undersampled F090W
images, the low IVM weights of the central pixels of the PSF have now
disappeared. This is reflected in the integrated flux measurements of the star
in the resampled images, which are now consistent with those in the flat-fielded
images to within 0.5\% for both dither positions (see
Figure~\ref{fig:cal_i2d_F480M_after}).  

\section{Summary}
\label{s:summ}
We describe the negative impacts of the Brighter-Fatter Effect (BFE) to data of
the NIRISS instrument aboard the James Webb Space Telescope (JWST). The efficacy
of the BFE becomes significant when a pixel's accumulated signal levels reaches
beyond a certain threshold (of order 20,000\,--\,25,000 ADU depending on the sharpness
of the point spread function (PSF)), at which point charge starts to migrate to
neighboring pixels with lower signal levels by detectable amounts.   
This process leads to detections of ``jumps'' in reads within integration ramps of
the affected pixels by the JWST calibration pipeline. The reads that get flagged
as jumps for the peak pixel due to this effect are typically different from those
flagged for the neighboring pixels, causing incorrect determinations of both
peak pixel count rate and total source signal. Furthermore, the jump flags caused
by this effect (which can be significant in number) cause low weights for the central
pixels of PSFs of bright stars in the ``inverse variance mapping'' (IVM)
weighting scheme, which is the default scheme used in the {\tt resample} step of
the JWST calibration pipeline which resamples images onto a distortion-free pixel
grid. These low IVM weights for the bright central pixels can cause significant
loss of source flux in the output images of the {\tt resample} step relative to
the input images, especially in the case of spatially undersampled
images. Flux losses of $>$\,50\% have been identified in this context.

We describe an algorithm to mitigate the effects mentioned above, called the {\tt
  charge\_migration} step, which has been implemented within the {\tt 
  calwebb\_detector1} stage of build 10.0 of the JWST calibration pipeline,
which was released on December 5, 2023. This step limits the negative impacts of
the BFE in NIRISS imaging data to within 1\% in signal level, both for
flatfielded images and images resampled to a distortion-free pixel grid. 

\acknowledgments

The data presented in this paper were obtained from the Mikulski Archive for
Space Telescopes (MAST) at the Space Telescope Science Institute. The specific
observations analyzed can be accessed via
\dataset[https://doi.org/10.17909/8jct-sx76]{https://doi.org/10.17909/8jct-sx76}.
STScI is operated by the Association of Universities for Research in Astronomy,
Inc., under NASA contract NAS5–26555. Support to MAST for these data is provided
by the NASA Office of Space Science via grant NAG5–7584 and by other grants and 
contracts.  
We acknowledge the efforts of Eddie Bergeron (STScI) during his early
investigations of the Brighter-Fatter Effect across the detectors of
the various JWST instruments. We thank the referee for their useful
comments and questions that helped improve the clarity of the text. 
This research has made use of NASA’s Astrophysics Data System.


\vspace{2mm}
\facilities{JWST (NIRISS)}


\software{{\tt Python} \citep{vanrossum+09}, {\tt AstroPy}
  \citep{astropy13,astropy18,astropy22}, {\tt matplotlib}
  \citep{matplotlib07}, {\tt NumPy} \citep{numpy20}, {\tt Photutils}
  \citep{bradley+22}}  



\clearpage

\appendix

\section{Determination of \texttt{signal\_threshold} for NIRISS imaging data}
\label{s:thresholds}
To determine appropriate values for the {\tt signal\_threshold} parameter in the
context of the {\tt charge\_migration} step, the goal we aim to achieve is a
situation where integrated fluxes of (bright but unsaturated) stars measured from 
resampled {\tt \_i2d} images are consistent with those measured from individual
flatfielded {\tt \_cal} images to within 1\%. From the results described in
Sections~\ref{s:impact} and \ref{s:algorithm}, this corresponds to ramp slopes
calculated for the peak pixel of the PSF that are consistent to within 1\% with
those determined from the groups with signal levels low enough for the efficacy 
of BFE to be negligible. 

With this goal in mind, we identify suitable imaging datasets for a
variety of NIRISS passbands. These images contain a star for which the
integrations of the peak pixel feature the following:
\begin{itemize}
 \item  The integrations contain at least 3 groups with accumulated signal level $<$
  18,000 ADU (at which no sign of charge migration has been detected). This is
  to assure a robust ramp slope measurement. For the discussion below, we
  define $N_{\rm 18K}$ as the last group in the integration ramp for which the peak
  pixel reaches a signal level $<$ 18,000 ADU.
 \item The integration ramps contain at least 2 groups with accumulated signal
   level $>$ 25,000 ADU. This is to ensure a robust quantification of the
   effect of charge migration on the derived ramp slope.
\end{itemize}
Exposures with these features identified among publically available NIRISS
data are listed in Table~\ref{tab:thresholds}. For each of these exposures, we
obtain linearized ramps by running the {\tt calwebb\_detector1} pipeline with
the {\tt save\_calibrated\_ramp = True} setting, but \emph{without} the new
  \texttt{charge\_migration} step activated. 
Ramp slopes {\it slope}$_N$ are then calculated for both the peak pixel
and the sum of the inner 5$\times$5 pixels for groups $N$ with
$N_{\rm 18K} \leq N \leq {\rm NGROUPS}$. Note that the calculation of {\it
  slope}$_N$ involves groups 1 through $N$, using linear regression, and
we ignore jump detections in the slope determination in this case.   
We define the ``Fractional Count Rate'' {\it fracrate}$_N$ as 
\begin{equation}
  {\it fracrate}_N = {\it slope}_N / {\it slope}_{N_{\rm 18K}}
  \label{eq:fracrate}
\end{equation}
We then calculate {\it fracrate}$_N$ for each group $N > N_{\rm 18K}$ up the
ramp, averaging over all ramps in the exposure using iterative $3 \sigma$
clipping statistics. Finally, we use linear interpolation to calculate the
signal levels for which {\it fracrate}$_N$ equals 0.99 and 0.98 for the peak
pixel, i.e., the signal levels at which the BFE has caused the ramp
slope (or derived count rate) of the peak pixel to decrease by 1\% and 2\%,
respectively. 
For the remainder of this Appendix, we define these two signal levels as
$S_{0.99}$ and $S_{0.98}$, respectively.

\begin{figure*}[htb]
  \centerline{\includegraphics[width=15.5cm]{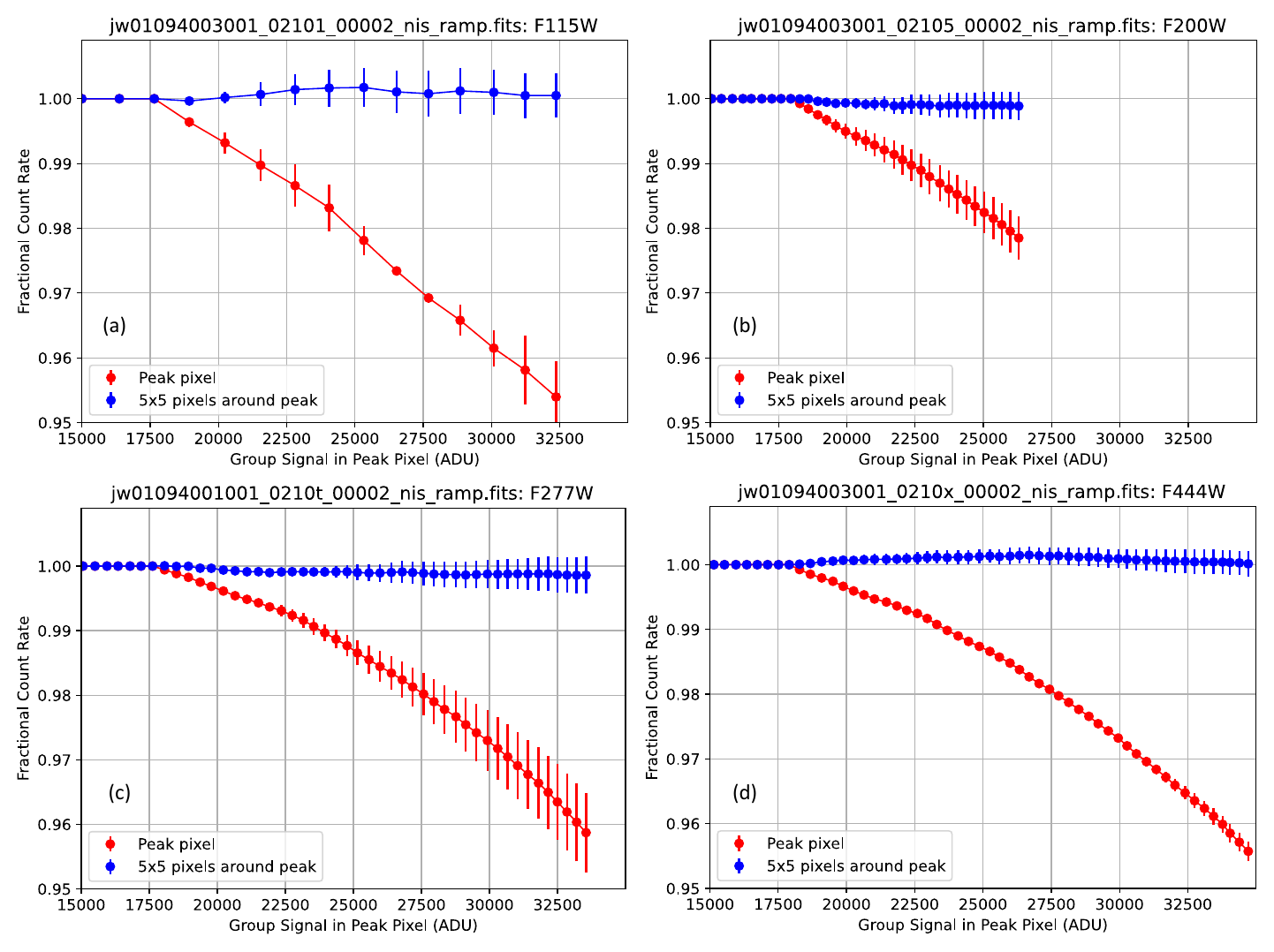}}
  \caption{{\it Panel (a)}: Fractional Count Rate (see eq.\ \ref{eq:fracrate})
    as a function of peak pixel signal level attained during the integration
    ramps for F115W exposure {\tt jw01094003001\_02101\_00002}. Red symbols
    indicate data for the peak pixel while blue symbols indicate data for the set
    of the inner 5$\times$5 pixels of the PSF. See Appendix~\ref{s:thresholds} for 
    discussion.  
    {\it Panel (b)}: same as panel (a), but now for F200W exposure {\tt
      jw01094003001\_02105\_00002}. 
    {\it Panel (c)}: same as panel (a), but now for F277W exposure {\tt
      jw01094001001\_0210t\_00002}. 
    {\it Panel (d)}: same as panel (a), but now for F444W exposure {\tt
      jw01094003001\_0210x\_00002}. 
    \label{fig:fracrates}
  }
\end{figure*}

Values of {\it fracrate}$_N$ as a function of signal level at group $N$ are
plotted in Figure~\ref{fig:fracrates} for four different filter
passbands. Note that {\it fracrate}$_N$ steadily decreases beyond
$\sim$\,18,000 ADU for the peak pixel, while it stays constant to well within
1\% for the set of the inner 5$\times$5 pixels. This illustrates that even
  without application of the \texttt{charge\_migration} algorithm presented in this
  paper, charge migration caused by BFE does in principle not impact \emph{total} flux
measurements of (non-saturated) stars provided that (1) the measurement aperture
radius is large enough (i.e., $\ga$ 2 pixels) and (2) jump detections due
  to the BFE are dealt with properly during ramp slope fitting.
Application of the \texttt{charge\_migration} step mitigates the negative  
  impacts of the BFE to the count rate levels derived by ramp fitting for the
  inner few pixels of bright stars or other unresolved sources. 
Figure~\ref{fig:fracrates} also suggests that the functional form of
  \textit{fracrate}$_N$ for the peak pixel is consistent across different
  datasets and that it may be possible to model BFE as functions of signal level and
  contrast with neighboring pixels in NIR data simulation software for JWST such as
  \texttt{MIRAGE}\footnote{\url{https://mirage-data-simulator.readthedocs.io/en/latest}}. This
  will be further explored in the future. 

\begin{figure}[hbp]
  \centerline{\includegraphics[width=8.3cm]{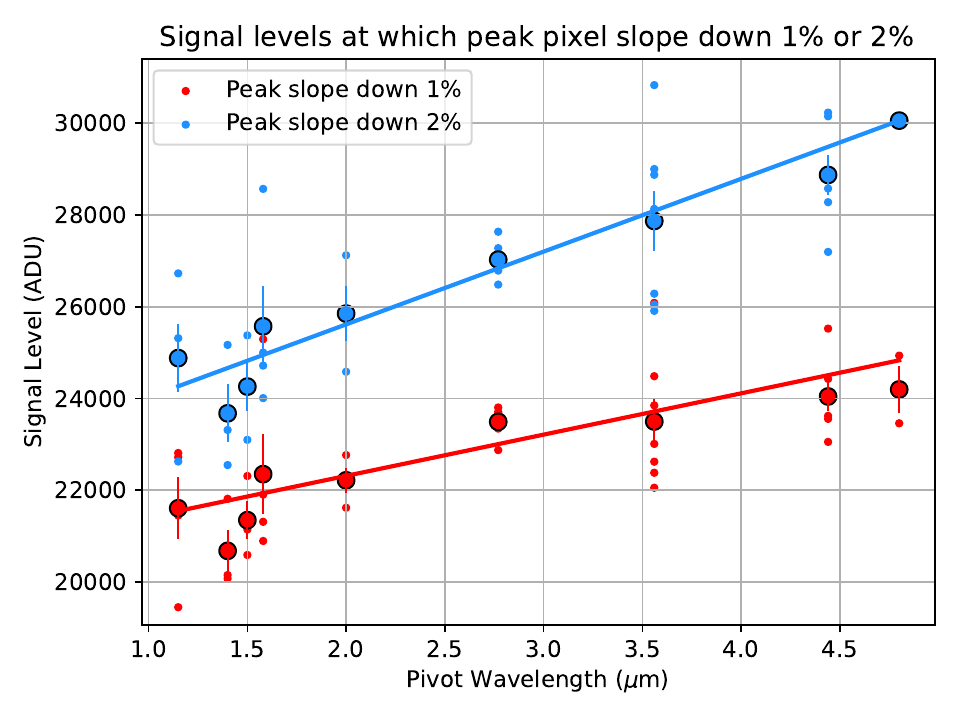}}
  \caption{Signal levels $S_{0.99}$ (red circles) and $S_{0.98}$ (blue circles)
    versus filter pivot wavelength. Small circles represent measurements of
    individual stars while large circles with black rims and errorbars
    represent average values and standard deviations of the measurements for a
    given filter. Linear fits to the average values of $S_{0.99}$ and
    $S_{0.98}$ as function of pivot wavelength are shown as solid lines with
    the respective line colors. 
    See Appendix~\ref{s:thresholds} for discussion.  
    \label{fig:thresholds}
  }
\end{figure}

Values of $S_{0.99}$ and $S_{0.98}$ for all exposures listed in
Table~\ref{tab:thresholds} are included in that table and plotted 
as a function of filter pivot wavelength in Figure~\ref{fig:thresholds}. There
is a significant dependence on filter pivot wavelength, which we attribute to
the known increase of BFE efficacy with increasing pixel-to-pixel contrast
\citep[e.g.,][]{hirachoi20} 
since this contrast increases in narrower PSFs, i.e., those with decreasing
pivot wavelengths. Quantitatively, linear least-squares fits to the 
values of $S_{0.99}$ and $S_{0.98}$ as a function of pivot wavelength
$\lambda_P$ in $\mu$m yield the following results:
\begin{eqnarray}
  S_{0.99} & \; = \; & 20514 \; (\pm\, 2.7\%) \; + \; 900 \; (\pm\, 22\%) \; \lambda_P
   \label{eq:S0p99} \\
  S_{0.98} & \; = \; & 22443 \; (\pm\, 1.4\%) \; + \; 1586 \; (\pm\, 4.2\%) \; \lambda_P 
  \label{eq:S0p98}
\end{eqnarray}

Finally, values for the {\tt signal\_threshold} parameter of the {\tt charge\_migration}
step were chosen according to eq.\ \ref{eq:S0p99}.  These values have been
implemented as parameter reference files in the JWST Calibration Reference Data
System \href{https://jwst-crds.stsci.edu}{(CRDS)} (context \#\,1135) 
as of the release of version 10.0 of the Operational JWST Calibration
Pipeline. As such, the {\tt charge\_migration} pipeline step will be
automatically applied to NIRISS AMI, imaging, and WFSS data taken after December
5, 2023\footnote{Updates to the parameter reference files for the AMI modes with
the non-redundant mask in the pupil wheel are in process and planned to be
implemented before this paper is published.}. 
NIRISS data taken before that date will be recalibrated
over the few weeks after that date and made available again in the MAST
archive. In the mean time, users can download the JWST python
package version 1.12.3 (or higher) using pip\footnote{see
\url{https://github.com/spacetelescope/jwst}} in order to run the {\tt
  charge\_migration} package on NIRISS data that was downloaded before
December 5, 2023. 

\begin{table*}[p]
 \caption{Datasets used to determine {\tt signal\_threshold} for NIRISS
   imaging and WFSS \label{tab:thresholds}}
 \footnotesize
 \vspace*{-2ex}
\begin{center}
 \begin{tabular}{@{}lrcrrlcc@{}} \tableline \tableline
   JWST Dataset Name & Observing Date & Filter & NGROUPS & NINTS & Subarray & 
   $S_{0.99}$ & $S_{0.98}$  \\ 
   \multicolumn{1}{c}{(1)} & \multicolumn{1}{c}{(2)} & \multicolumn{1}{c}{(3)} &
   \multicolumn{1}{c}{(4)} & \multicolumn{1}{c}{(5)} & \multicolumn{1}{c}{(6)} &
   \multicolumn{1}{c}{(7)} & \multicolumn{1}{c}{(8)} \\ [0.5ex] \tableline
 & &  \\ [-2.8ex]  
jw01094001001\_02101\_00002 &  May 5, 2022 & F115W &  25 & 5 &  SUB64 & 22720 & 26726 \\
jw01094001001\_02103\_00002 &  May 5, 2022 & F150W &  40 & 5 &  SUB64 & 22311 & 25375 \\
jw01094001001\_02105\_00002 &  May 5, 2022 & F200W &  75 & 5 &  SUB64 & 21619 & 24583 \\
jw01094001001\_0210b\_00002 &  May 5, 2022 & F140M &  20 & 3 & SUB128 & 20074 & 22549 \\
jw01094001001\_0210d\_00001 &  May 5, 2022 & F158M &  25 & 3 & SUB128 & 20895 & 24008 \\
jw01094001001\_0210d\_00002 &  May 5, 2022 & F158M &  25 & 3 & SUB128 & 21313 & 24700 \\
jw01094001001\_0210f\_00002 &  May 5, 2022 & F115W &   7 & 3 & SUB128 & 22812 & 25314 \\
jw01094001001\_0210p\_00002 &  May 5, 2022 & F356W & 180 & 3 & SUB128 & 22055 & 26047 \\
jw01094001001\_0210p\_00001 &  May 5, 2022 & F356W & 180 & 3 & SUB128 & 22382 & 25910 \\
jw01094001001\_0210r\_00001 &  May 5, 2022 & F444W & 340 & 3 & SUB128 & 23054 & 27194 \\
jw01094001001\_0210r\_00002 &  May 5, 2022 & F444W & 340 & 3 & SUB128 & 23621 & 28575 \\
jw01094001001\_0210t\_00001 &  May 5, 2022 & F277W &  80 & 3 & SUB128 & 22875 & 26482 \\
jw01094001001\_0210t\_00002 &  May 5, 2022 & F277W &  80 & 3 & SUB128 & 23804 & 27634 \\
jw01094001001\_0210v\_00001 &  May 5, 2022 & F277W &  20 & 3 & SUB256 & 23708 & 27011 \\
jw01094001001\_0210v\_00002 &  May 5, 2022 & F277W &  20 & 3 & SUB256 & 23671 & 26786 \\
jw01094001001\_0210x\_00001 &  May 5, 2022 & F444W &  90 & 3 & SUB256 & 23556 & 30230 \\
jw01094001001\_0210x\_00002 &  May 5, 2022 & F444W &  90 & 3 & SUB256 & 24427 & 28803 \\
jw01094001001\_0210z\_00001 &  May 5, 2022 & F356W &  40 & 3 & SUB256 & 24485 & 29008 \\
jw01094001001\_0210z\_00002 &  May 5, 2022 & F356W &  40 & 3 & SUB256 & 23849 & 28132 \\
jw01094001001\_0211b\_00002 &  May 5, 2022 & F140M &   6 & 3 & SUB256 & 20152 & 23314 \\
jw01096001001\_02101\_00002 & May 14, 2022 & F200W &   8 & 1 &   FULL & 22767 & 27121 \\
jw01096001001\_02107\_00001 & May 14, 2022 & F158M &  14 & 1 &   FULL & 25292 & 28567 \\
jw01096001001\_0210d\_00001 & May 14, 2022 & F277W &  12 & 1 &   FULL & 23343 & 26963 \\
jw01096001001\_0210h\_00001 & May 14, 2022 & F356W &  18 & 1 &   FULL & 26084 & 30829 \\
jw01096001001\_0210h\_00002 & May 14, 2022 & F356W &  18 & 1 &   FULL & 23010 & 28875 \\
jw01093023001\_03103\_00001 & June 5, 2022 & F480M &  13 & 232 & SUB80 & 24934 & 30043 \\
jw01093023001\_03103\_00002 & June 5, 2022 & F480M &  13 & 232 & SUB80 & 23459 & 30068 \\
jw01094003001\_02101\_00002 & June 6, 2022 & F115W &  25 & 5 &  SUB64 & 21449 & 24853 \\
jw01094003001\_02103\_00002 & June 6, 2022 & F150W &  40 & 5 &  SUB64 & 21144 & 24305 \\
jw01094003001\_02105\_00002 & June 6, 2022 & F200W &  75 & 5 &  SUB64 & 22263 & 25846 \\
jw01094003001\_02109\_00002 & June 6, 2022 & F150W &  10 & 3 & SUB128 & 20590 & 23099 \\
jw01094003001\_0210r\_00001 & June 6, 2022 & F444W & 340 & 3 & SUB128 & 25523 & 30150 \\
jw01094003001\_0210r\_00002 & June 6, 2022 & F444W & 340 & 3 & SUB128 & 24080 & 28278 \\
jw01094003001\_0210b\_00002 & June 6, 2022 & F140M &  20 & 3 & SUB128 & 21814 & 25167 \\
jw01094003001\_0210d\_00001 & June 6, 2022 & F158M &  25 & 3 & SUB128 & 21905 & 25010 \\
jw01094003001\_0210f\_00002 & June 6, 2022 & F115W &   7 & 3 & SUB128 & 19451 & 22628 \\
jw01094003001\_0210p\_00002 & June 6, 2022 & F356W & 180 & 3 & SUB128 & 22623 & 26284 \\
jw01094003001\_0210t\_00002 & June 6, 2022 & F277W &  80 & 3 & SUB128 & 23544 & 27275 \\  [0.5ex] \tableline
\end{tabular}
\end{center}
\tablecomments{Column (1): dataset name. Column (2): observing date. Column (3):
  imaging filter. Column (4): number of groups per integration ramp. Column (5):
  number of integration ramps. Column (6): subarray used. Column (7): value of
  $S_{0.99}$ (see text in Appendix~\ref{s:thresholds} for its definition). Column
  (8): value of $S_{0.98}$. }
\end{table*}





\clearpage
\bibliography{PGrefs}{}
\bibliographystyle{aasjournal}

\end{document}